# Asymmetrical inheritance of plasmids depends on dynamic cellular geometry and volume exclusion effects

Jai A. Denton[1], Atiyo Ghosh[1] and Tatiana T. Marquez-Lago*

Integrative Systems Biology Unit, Okinawa Institute of Science and Technology, Onna-son, Okinawa 904-0495, Japan.

[1] Both authors contributed equally to the work and are presented in alphabetical order.

* Correspondence: tatiana.marquezlago@gmail..com

**Abstract**

The asymmetrical inheritance of plasmid DNA, as well as other cellular components, has been shown to be involved in replicative aging. In *Saccharomyces cerevisiae*, there is an ongoing debate regarding the mechanisms underlying this important asymmetry. Currently proposed models suggest it is established via diffusion, but differ on whether a diffusion barrier is necessary or not. However, no study so far incorporated key aspects to segregation, such as dynamic morphology changes throughout anaphase or plasmids size. Here, we determine the distinct effects and contributions of individual cellular variability, plasmid volume and moving boundaries in the asymmetric segregation of plasmids. We do this by measuring cellular nuclear geometries and plasmid diffusion rates with confocal microscopy, subsequently incorporating this data into a growing domain stochastic spatial simulator. Our modelling and simulations confirms that plasmid asymmetrical inheritance does not require an active barrier to diffusion, and provides a full analysis on plasmid size effects.

**Author Summary**

Cellular division is a highly controlled critical process on which all life depends. Among many other organisms, baker's yeast (*Saccharomyces cerevisiae*) cells undergo asymmetrical inheritance. As the name suggests, yeast daughter cells inherit more of some cellular components, and less of others, than expected by cellular volume alone. While this process has been shown to be important in reducing the effects of cellular aging in daughter cells, the mechanisms behind this asymmetrical inheritance are still not well understood. It has been proposed that asymmetrical inheritance is achieved through the dynamical geometry of the cell alone, additional diffusion barriers, or a combination of the two. These contrasting views have created a controversy in the field of asymmetric inheritance of nucleoplasmic components. In this study we model the inheritance of a yeast plasmids inside the nucleus by using cell geometries and diffusion rates derived from confocal microscopy data into a new growing domain spatial stochastic simulator. We confirm that strong asymmetrical retention of the plasmid in the mother cell can be achieved without the need for a distinct barrier, and can solely rely on the moving boundaries and cellular geometry of the dividing cell.

**Introduction**

Organisms need to properly manage cellular aging processes to ensure population survival. Complex multicellular organisms do so by following a number of strategies, including the separation of a germ line early in development or the maintenance of pluripotent stem cells. Many strategies are not available to single-cell organisms, but they efficiently reduce the impact of aging through asymmetric cell division, wherein cellular components are not divided equally between mother and daughter cells during mitosis. This allows for titrating away cellular components such as extra chromosomal DNA, aged organelles and mis-folded proteins that could potentially shorten the life of subsequent cell generations.

For *Saccharomyces cerevisiae*, there are a limited number of divisions an individual cell can undergo before becoming senescent [1] . This process of replicative aging is undeniably complex, with many environmental and genetic factors influencing it. However, recent studies have demonstrated that asymmetrical inheritance of cellular components, where cellular components tend to remain disproportionally in the mother cell, are a key factor in yeast aging. Numerous experiments have demonstrated that the asymmetrical inheritance of mitochondria [2] and plasmids, such as autonomously replicating extrachromosomal rDNA circles (ERCs) [3] and autonomously replicating ARS domain containing plasmids (ARS plasmids) [4] , are linked to replicative aging (cf. reviews in [5–7] ).

There are notable differences between ERCs and ARS plasmids. Although both are self-replicating pieces of circular DNA, ERCs arise spontaneously from the resolution of Holiday junctions and appear ubiquitous throughout *Eukaryota* (reviewed in [5,6] ). In *S. cerevisiae,* ERCs arise from the 9.1kb locus containing ~150 repeats of the rRNA precursors [3] . These plasmids accumulate in yeast cells over their life and can reach >500 copies [3] . In contrast, ARS plasmids are not naturally occurring, but are the result of modifying the naturally occurring 2µm plasmid [8] . In comparison to ERCs, the ARS plasmids tend to have lower copy numbers in the range of 10-100 [9] .

Although the existence of circular DNA asymmetrical inheritance has been known for a long time, there are still considerable gaps in the understanding of the mechanisms underlying this asymmetry. The current debate, regarding a potential mechanism for this asymmetrical inheritance, has two main postures: (1) either a morpho-kinetic passive diffusion model, with asymmetry being generated due to time and cell morphology [10–12] ; or (2) a barrier division model, where physical agents such as cellular machinery actively restrain DNA transmission [13–15] .

Many contemporary studies of plasmid inheritance have focused on only a single plasmid type [10,13–15] . The lack of direct comparisons between ERCs or ARS plasmids compounds the difficulty in comparing these studies. Thus, it becomes necessary to highlight several factors that may be crucial to understanding the differences between observations and studies. First, *S. cerevisiae* mutants with extended mitosis length show a lower level of plasmid asymmetrical inheritance [10] . This can be intuitively expected, given lengthier time frames effectively allow for diffusion of more cellular components than shorter time frames. More recently, it was reported that ERCs are bound to the nuclear pore complex via interaction with the master regulator complex SAGA [13] . The authors of [13] also note that, even in the absence of nuclear pore binding, passive diffusion continues to play a critical role in asymmetrical inheritance, attributing this effect to a potential diffusion barrier. However, the movement of nuclear pores from the mother cell to the daughter themselves further complicates the diffusion of nuclear pore fused plasmids [16–18] . For instance, by using photo-switchable florescence microscopy, it was shown that nuclear pores effectively diffuse from the mother to the daughter nuclear lobe during anaphase [16] . Moreover, there is evidence for the active transport, mediated by the protein Nsp1, of the nuclear pores from mother to daughter [18] .

In addition, computational models have suggested that moving boundaries can have an important effect [10] . However, these conclusions were derived by comparing the equilibration time of diffusing particles in static geometries to the overall length of mitosis. The question remains as to the magnitude of this effect, which can only be investigated by building a framework where moving boundaries are explicitly modelled. By the same token, most previous studies have focussed on treating plasmids as point particles [10,15] , rather than particles with volume, as is actually the case. Given that the region between daughter and mother nuclei can narrow in late anaphase, it stands to reason that the role of plasmid size in driving asymmetric cell division is understated in all existing models. Lastly, there might be large heterogeneity within nuclear division profiles, which in turn could lead to heterogeneous transmission rates.

The effects of particle size, moving cell boundaries and cell-to-cell variability, on plasmid diffusion, have remained largely unstudied. In this paper, we aim to incorporate these factors into the morpho-kinetic model, by using mathematical and computational modelling based on cell geometries and plasmid motion as obtained from confocal microscopy. We have used measured *S. cerevisiae* cellular geometries as a basis for modelling the geometry within which the inheritance of a single plasmid happens. In addition we tracked individual plasmids to compare experimental data with these simulation results. By incorporating dynamical changes to cell morphology, explicitly into a spatial stochastic simulation framework, and plasmid diffusion rates, we explore the opposing postures of the debate from a quantitative perspective.

## Results

Effect of Moving Boundaries and Heterogeneity

Confocal microscopy showed a heterogeneous profile of nuclear geometries over the course of anaphase. Fig. 1 shows 10 of these arbitrarily chosen anaphase geometries, and their idealisation *in silico* as per the previously described method. The final proportion of the nucleus occupied by the daughter cell, in these 10 measured cells, was estimated to be in the range of 33-48%, with a mean of 42%. Plasmid diffusion rates in individual cells were found to be in the range of 0.0010-0.0046 $\mu m^2/s$, with a mean of 0.0027 $\mu m^2/s$, consistent with previously published studies [10]. Numerical simulations with 300 independent repeats were performed with this diffusion coefficient in the 10 measured geometries. These simulations were further repeated for two plasmid representations: point particles, and 50 nm solid volume spheres, the latter of which accounts for volume exclusion effects explicitly. Fig. 2 shows the proportion of plasmids contained within the daughter nucleus over the course of anaphase, superimposed onto the proportion of the daughter nucleus by volume over the course of anaphase. If the diffusion of plasmids within moving boundaries leads to a homogeneous distribution in space, one would expect the proportion of plasmids in the daughter to coincide with the corresponding proportion of volume occupied by the daughter lobe. However, we observed plasmids being actually retained in the mother, as also evidenced by the difference between the proportions in Fig. 2. This effect was solely produced by the dynamic changes of cell geometry, as there was no diffusion barrier incorporated in the study, as previously done for nuclear pores, nucleoplasmic and nuclear membrane proteins in [15] . Simulations of point particles showed that the final transmission of plasmids to the daughter nucleus varied from a minimum 12.1 +/- 3.8% (Fig. 2b) to a maximum of 26.2 +/- 5.1% (Fig. 2g (with 95% confidence intervals), with a mean of 21.4%. Representing the plasmid as a 50 nm sphere altered these percentages to 3.2 +/- 2.0% (Fig. 2f) to a maximum of 20.0 +/- 4.6% (Fig. 2j), with a mean of 13.4%. In all simulations of point particles and particles with 50 nm radius, plasmid transmission to the daughter nucleus was found to be significantly less that the proportion of volume represented by the daughter nucleus at a 95% confidence interval. These ranges are consistent with expectations from experimental studies [10,14] . However, these specific experimental measures require further thought, as will be explained in the Discussion section.

Effect of Diffusion Rate

Numerical simulations over the measured geometries were executed for diffusion constants in the range of 0.001-0.009 $\mu m^2/s$, such that this range encompassed the empirically determined range of 0.0010-0.0046 $\mu m^2/s$. For each diffusion constant, we further investigated a range of plasmid sizes, ranging from point particles to spheres with 100 nm radii. Fig. 3 shows the effect of changing diffusion constants on the final percentage of plasmids transmitted to the daughter. For each parameter set, simulations of single plasmids in each geometry were executed with 1000 repeats. As can be seen in Fig. 3k, the averaged effect of increasing diffusion constants tends to increase the transmission of plasmids to the daughter lobe. However, investigations in different geometries reveal further trends. For instance, in Fig. 3f and 3g, we observe insignificant effects of increasing the diffusion rate on the transmission of 50 nm plasmids, whereas transmission of point particles show more sensitivity to diffusion constant values. This is consistent with the nuclear geometry profile of these cells (see Fig. 1, Cell 6 and 7) which show that these cells establish an especially narrow bridge early on and maintain it throughout anaphase, thus inhibiting transmission of sizeable plasmids through the bridge from an early stage irrespective of diffusion rates, which is also consistent with the results in Fig. 2f and 2g. In Fig. 3a and 3d, we observe that increasing the diffusion coefficient leads to a larger increase in plasmid transmission as compared to other cells. This observation is consistent with this cell having a wider bridge during anaphase (see Fig. 1, Cell 1 and Cell 4).

Effect of Plasmid Size

A parameter sweep covering biologically feasible representations of plasmid sizes, as shown in Fig. 4, reaffirms conclusions from previous sections. The effect of increasing plasmid volume generally decreases the transmission rate to the daughter nucleus. However, we also observe geometry specific effects. Fig. 4f and 4g show especially large effects of plasmid size, consistent with their geometries forming narrow necks at the start of anaphase (as in Fig. 1, Cell 6 and 7).

These simulation results suggested that changing plasmid size should effect transmission of the plasmid from mother to daughter, in turn affecting the plasmid loss rate. The pAA4 plasmid contains 256 tandem *lacO* binding sites thereby allowing the reporter protein, LacI fused to GFP, to detect their cellular localisation. Thus, when a strain containing both the pAA4 plasmid and the GFP fusion is grown, the numerous GFP fusion molecules will bind the plasmid. The resulting plasmid with GFP bound could have either a more relaxed or compact configuration. The mitotic stability, the transmission percentage of a plasmid in selective conditions, and the plasmid loss rate, the rate at which a dividing cell produces a plasmid free daughter, was calculated for two strains containing the pAA4 plasmid: GA180, a wild type strain, and GA1320, GA180 containing the GFP-LacI reporter and NUP49-GFP. We found no statistically significant difference between mitotic stability between the two strains, but the plasmid loss rate was significantly lower in the GFP strain (see Table 1). The latter suggests a more compact configuration of pAA4 with GFP bound.

**Table 1** – Plasmid Loss Rate. A star * denotes statistically significant differences, with $P < 0.01$. Mitotic stability was not significantly different between the WT and the GFP strain ($P = 0.21$). However, the plasmid loss rate was significantly different ($P = 0.0048$).

|  | Mitotic Stability | | Plasmid Loss Rate | |
| --- | --- | --- | --- | --- |
| **Strain** | Average | StDev | Average | StDev |
| Wild Type (GA180) | 25.3% | 5.2% | 0.37* | 0.09 |
| GFP Fusion (GA1320) | 18.9% | 5.4% | 0.17* | 0.09 |

**Discussion**

The work here demonstrates that heterogeneity, in terms of cell morphology within a *S. cerevisiae* population, greatly influences the inheritance of individual autonomously replicating plasmids. Moreover, the asymmetry of this inheritance is also influenced by the moving boundaries of the cell, the rate of diffusion and the size of the diffusing plasmid / particle.

The effect of moving boundaries is best exemplified in Fig. 3. In regions of parameter space where we expect moving boundaries to have a minimal effect (i.e. where plasmids are modelled as points with high diffusion coefficients), numerical simulations demonstrate an asymmetric distribution of plasmids towards the mother. Moreover, this effect was observed in all measured cell geometries. For our measured geometries, particles diffusing with speeds of $D = 0.001 - 0.009 \mu m^2 / s$ were found to be sensitive to changes in diffusion constant. This is in line with previously reported values [12,13].

It is worth mentioning the work in [12] incorporates the effects of moving boundaries analytically, by using a narrow-escape approximation to calculate the distribution of escape times from the mother to a daughter. Here, we additionally considered several empirically-determined geometries, and included plasmid size into our investigations, complementing and extending the results in [12]. Further work in generalising the analytical techniques in [12] to encompass these factors could lead to interesting further work.

Owing to the size, behaviour, compartmentalisation and function of the numerous cellular components, it is unlikely there is a single mechanism driving asymmetrical inheritance of these cellular components. The analytical model of diffusion-driven transfer in [12] demonstrated there is a threshold passive diffusion rate with which cellular components remain within the mother cell during cell division. Therefore, smaller particles, moving at speeds near or faster than this threshold, may require additional mechanisms to remain in the mother cell during division. This does not preclude the existence of barriers, such as the one hypothesized in [15], but there are as yet few concrete suggestions as to what might constitute such barriers.

The effect of cell heterogeneity, diffusion coefficients and particle size were all found to be mutually dependent, as shown in Fig. 3 and 4. Generally, increasing plasmid size and decreasing diffusion coefficients decreased plasmid transmission to the daughter lobe. This is consistent with previous empirical studies that linked increasing diffusion constants with increasing plasmid inheritance [19]. However, there were geometry-specific exceptions. Particularly, for biologically plausible plasmid sizes, there were measured nuclear geometries where increasing diffusion coefficients had no meaningful effect on the final transmission of plasmids (Fig. 3f and 3g). We anticipate that in these scenarios, the geometric constraints were limiting factors in the transmission of plasmids, since these cells formed especially narrow bridges between the daughter and mother lobes in early anaphase. This is consistent with previous studies, namely those with static boundaries that explored the manipulation of the nuclear bridge width [10]. However, there were also cells that showed stronger sensitivity to increasing diffusion coefficients (e.g. Fig. 3d). These cells had a wider bridge between the mother and daughter lobes throughout anaphase. Thus, these findings caution against using idealised geometries for such studies. However, it should be emphasised that yeast mitosis and cellular geometries are very variable, heavily influenced by numerous factors including growth media, temperature and genetic background. As such, our aim of simulating numerous measured nuclear geometries was to demonstrate that this variation in mitosis could have an important effect on plasmid segregation. Furthermore, there is a great sensitivity of transmission rates to plasmid sizes over biologically relevant parameters (see Fig. 4). These findings were reinforced empirically by the observation that loss rates of plasmids bound with GFP were significantly lower than plasmids with no GFP bound.

For numerical simulations, we considered plasmids as hard spheres, as opposed to the biological reality of them being flexible polymers. However, we anticipate that our representation of hard spheres helps to establish upper and lower bounds of the transmission of such a polymer, which should intuitively lie somewhere between that of a point particle and a large solid sphere. Thus, we anticipate that our findings will hold if simulation methodologies allowing for more biologically faithful plasmid morphologies were introduced, where plasmid volumes vary in time.

There remain questions on how theoretical studies can be compared to experimental results. Current studies argue for asymmetric distribution of plasmids via loss rates, which is the proportion of plasmids lost from a population per generation under non-selective conditions. In order to build a bridge from the microscopic level of the cell to the macroscopic level of loss rates, more information concerning copy numbers of ARS plasmids and their replication rates are essential. There remain difficulties in extrapolating the meaning of empirically determined loss rates to transmission rates of plasmids.

Presently there are three methods for measuring plasmid loss rate: 1) a pedigree based system that relies on cell dissections each generation [20] ; 2) a microscope approach that tracks plasmid movement from mother to daughter cell [13] ; and, 3) plasmid loss rates based on whole populations of cells and the likelihood of dividing cells to lose the plasmid each generation [10,21] . While each of these methods provide important insights into the asymmetrical inheritance of plasmids, questions need to be asked regarding the comparability of these methods. Influences of cell backgrounds, either strain or genotype, or the addition of fluorophores for plasmid tracking, may have important effects. These questions may make the direct comparison of experimental data between studies, using different methods, a complex process.

The comparison of studies is further complicated as presently analysis of circular DNA inheritance has tended to compare ERCs, ARS plasmids and synthetic DNA circles directly. However, there exists the possibility that the asymmetry in DNA circle inheritance is influenced by independent factors such as their numbers within a cell, their interactions with other members of their DNA class or differences in interactions with other cellular components. Furthermore, our results suggests that differences in diffusion coefficients and sizes of particles can have significant effects in inheritance, which compounds existing difficulties in comparing studies between different episomes. Thus, some of the conflict regarding asymmetrical inheritance of circular DNA may arise from the idea that there is a general barrier mechanism to circular DNA inheritance.

Our model does not preclude the existence of a diffusion boundary, such as the fusion of circular DNA molecules to the nuclear pore, but suggests that such barriers are not required to obtain asymmetrical inheritance if the current experimental predictions are accurate. Further work also needs to be done to determine if there is a general mechanism regulating the asymmetrical inheritance of DNA circles or if only some of the asymmetry is regulated through transmission barriers.

Lastly, our findings indicate that nuclear morphology during anaphase and episome size can have important effects in yeast replicative aging. Furthermore, we observe that cell-to-cell variability could lead to inhomogeneous aging rates among a yeast population. These factors are crucial considerations, given a better understanding of asymmetrical inheritance could potentially result in effective strategies to mitigate the effects of replicative aging.

**Materials & Methods**

Yeast Strains, Growth & Manipulations

The stains used in this study are summarized in Table 2. Yeast media was made as described in Sherman [22] , and *S. cerevisiae* strains were transformed using the lithium acetate method. Strains GA180 & GA1320 were transformed with the pAA4-*lacO* plasmid [23] . This plasmid contains a 256-tandom array of *lacO* binding sites that allow binding of a LacI-GFP fusion reporter protein.

**Table 2** – Yeast Strains used in this study.

| Name | Genotype | Publication |
|---|---|---|
| GA180 | *MATa ade2 trp1 his3, ura3 leu2 can1* | (W303 wild type) |
| GA1320 + pAA4 | GA180, *NUP49-GFP LacI-GFP, pAA4-lacO* | [24] |

Microscopy

For imaging, *S. cerevisiae* cells were grown in SC low fluorescence media overnight at 30°C and 400 µl of this culture was transferred to 5ml of fresh SC-URA-HIS low fluorescence media and grown for a further 4 hours at 30°C. Cells were immobilised on 0.7% low melting temperature agarose. The agarose squares were then inverted in 35mm µ-Dishes (Ibidi 81158). All microscopy was conducted at 30°C.

Cell Geometry

Cell geometry was visualised using a Zeiss LSM-780 microscope with a 100x oil emersion objective. GFP was visualised using a 488 wavelength Argon laser at 1% intensity. Two hundred 20 slice z-stacks, separated by 0.4µm, were taken at 30-second intervals. The FIJI installation of ImageJ was used for image analysis. Geometric coordinates were generated in ImageJ for growing cells, and nuclear lengths were generated accordingly (See Fig. 1 for geometry comparisons and S8 for geometry measurements). Ten individual cells were arbitrarily selected and imaged. These initial 10 nuclear geometries were used as the basis for the simulations. Subsequently, 50 additional cells were measured to examine how representative the original selection was. Although it is difficult for simple metrics to capture the complexity of dividing nuclei, several metrics were chosen and corresponding histograms were generated, with which we compared the initial selection to the subsequent 50 geometries (See Fig. S7). The metrics chosen aim to capture the size of both the mother and daughter cells and ratios between them. Based on this analysis our 10 cells are reasonably representative.

Plasmid Diffusion

Estimates of plasmid diffusion were obtained by using images generated on a Zeiss ELYRA PS.1 microscope with a 100x oil emersion objective. Plasmids in cells that contained a single bright spot were tracked. However, it is worth noting that these single plasmid spots may be made up of several plasmids (cf.

Plasmid Loss Rate). Two hundred time steps with six slice z-stacks, separated by 0.5μm, were taken at 5-second intervals. GFP was visualised using a 488-wavelength Argon laser at 1% intensity with 200ms exposure. The FIJI installation of ImageJ was used for image analysis and bleach correcting [25]. Each image was stabilised using the plugin Correct 3D Drift [26]. Individual cells were extracted from the image into new flies and movement of the plasmid was tracked using the plugin Manual Tracking with local barycentre correction. Tracking consisted of movement of the plasmid from time point to time point. However, tracking was halted when plasmid movement was localised to the nuclear membrane, and restarted once it moved away. This was due to technical limitation is discerning between the GFP tagged plasmids and the GFP tagged Nup49. Moreover, at 0.5μm, the z-slices were too far apart to accurately track plasmid movement through the z-axis. Therefore, plasmid movement was tracked on a single Z plane; when the plasmid changed Z position the tracking was stopped and restarted. This limits the data collected but reducing the distance between Z-slices resulted in greatly increased bleaching of GFP. Furthermore, assuming the diffusive motion is isotropic, the loss of resolution in the Z coordinate should lead to no difference in the estimated diffusion coefficient if the resulting data is treated as a two dimensional Brownian path. Between 50-100 plasmid jumps were recorded for each plasmid. A total of 25 plasmids were individually tracked from cells in both anaphase and interphase (See Supplementary Fig. 1 for microscopy examples). Individual diffusion constants were estimated per cell, chosen according to their stage in the cell cycle. For each cell, a distribution of 5-second jump sizes was calculated in a 2D plane. Assuming Brownian motion, we have that the diffusion coefficient, $D$, is then given by: $D = \frac{\langle x^2 \rangle}{2d\Delta t}$, where $\langle x^2 \rangle$ is the average jump size, $\Delta t$ is the time interval (5 seconds) and $d$ is the dimensionality (2 here, since we only track the movement in a plane). Particle jumps near nuclear membranes were ignored, since the input for simulations should be a free diffusion coefficient. The average obtained over all cells was 0.0027 $\mu m^2/s$. The average of interphase cells was 0.0025 $\mu m^2/s$; the average of anaphase cells was 0.0031 $\mu m^2/s$. Supplementary Fig. S2 shows histograms of jump size distributions calculated according to the Freedman-Diaconis method [27], while Supplementary Text S3 shows individual diffusion coefficients from each cell.

Plasmid Loss Rate

The plasmid loss rate was calculated based on the method described in Longtine et al., and used in Gehlen et al. [10,21]. Cultures of each strain were grown overnight in plasmid selective media (SC-URA). Dilutions of these cultures were grown on selective, SC-URA, and non-selective, SC, plates. The mitotic stability percentage, $T_0$, was calculated by dividing the colonies on selective media by those on non-selective media. These cultures were diluted 10-3 in non-selective media and grown for 21 hours. Dilutions of these cultures were then grown on selective, SC-URA, and non-selective, SC, plates. A quantity $T_1$ was calculated by dividing the colonies on selective media by those on non-selective media. The plasmid loss rate was calculated using the formula $1-\left(\frac{T_1}{T_o}\right)^{\frac{1}{n}}$, where *n* is the number of generations, derived from Longtine et al. [21].

A microscopy approach to plasmid loss was obstructed by plasmid clumping. Although the plasmids are often present in high numbers within a cell, there are numerous cells that have a single bright spot. This plasmid clumping behaviour has also been observed in previous studies [10,13]. It was found that in many of anaphase cells examined this single bright spot would split into two equally bright spots in late anaphase, with one entering the daughter.

*In Silico* Simulations

Diffusion of plasmids within dividing nuclei were simulated using tailor-made particle-tracking methods designed to incorporate moving boundary and plasmid size effects [28]. Nuclear geometries were approximated by two prolate spheroids connected by a cylindrical bridge, where the midpoint of the bridge was considered as the boundary between the mother and daughter nuclei. Relevant geometric parameters were measured by hand, one by one, from images obtained by confocal microscopy. For simulated times between confocal images, geometric parameters were linearly interpolated so as to allow for a smoothly changing simulated geometry (see Supplementary Video S4). The initial position of particles was sampled from a uniform distribution over the volume of the nucleus. At each time step, particles were propagated according to a Gaussian kernel. For particles propagated onto or outside the nuclear membrane, the particles were immediately remapped to their closest point within the nucleus. The numerical method is described in more detail in Supplementary Text S5 and [28] . For each simulation run, a single plasmid was simulated per cell. While we anticipate that there is significant potential for volume exclusion effects via multiple factors, e.g. via plasmid-chromosome, plasmid-RNA or plasmid-transcription factor interactions, such interactions cannot be investigated through currently available data. Thus we interpret these effects to be incorporated into the effective measured diffusion constants of plasmids. In this way, consistency is maintained between our observations from confocal microscopy and the simulated model system.

Selection of Simulation Time Step

Choosing a suitable time-step, $\delta t$, is of crucial importance. For simulations in free space, the mean-squared displacement provides a method by which we can choose a suitable time-step. For a particle with diffusion constant, $D$ diffusing freely in a $d$-dimensional Euclidean space, the mean-squared displacement is given by $\langle x^2 \rangle = 2dDt$, with the mean-squared displacement along a given direction given by $\langle x^2 \rangle = 2Dt$. Thus, for a desired spatial resolution, $s_{res}$, to be obtained from a simulation, this motivates the choice of time-step to be $\delta t = \frac{s_{res}^2}{2D}$, with the result that $s_{res} \propto \sqrt{\delta t}$ following. When considering the action of moving boundaries at small time scales, for most physical situations (including those we consider here, since we interpolate between geometries linearly), the movement of the boundary is linear with time. Consequently, any motion imparted on a diffusing particle in a short time scale by a moving boundary should be limited to $s_{bound} \propto \delta t$. Thus, provided time steps are small enough, the effect of moving boundaries on the spatial resolution of a particle simulation should be negligible relative to diffusive motion.

Here, to determine a suitable choice of $\delta t$, we considered the worst-case scenario for our simulation methodology: that of small diffusion constants, where the movement of boundaries is largest compared to that of diffusive motion. We considered a range of time-steps which were varied by a parameter $m$: $\delta t = \frac{s_{res}^2}{mD}$, with $s_{res} = 0.025 \mu m$ and $D = 0.001 \mu m^2 / s$ for point particles. The effect of changing $m$ on the final proportion of plasmids in the mother and daughter lobes was investigated. Simulations were repeated 5000 times, and geometries from Cell 1 were used. The results are shown in Fig. S6 with green dotted lines showing what is conventionally used in static geometries [29], and red dotted lines denoting the more stringent level that we use in this study to encompass boundary movement. Little effect of changing $m$ was noted in the results. For all other simulations in the manuscript, we use $m=4$ and $s_{res} = 0.025 \mu m$. Note that the resolution on our confocal microscopy images is approximately $0.16 \mu m$, thus our choice of spatial resolution is significantly finer than any geometric features that are measured.

**Acknowledgments**

The authors would like to thank Susan Gasser and the Gasser laboratory for providing strains and the pAA4 plasmid. Likewise, the authors would like to thank Zach Hensel for help and critical comments on the research results. The authors would also like to thank the two anonymous reviewers for their insightful comments.

**Author Contributions**

TML conceived the study. AG and JAD conducted the computational and microscopy work, respectively. TML, AG and JAD conducted analysis and wrote the manuscript.

**Funding information**

JAD, AG and TML were supported by subsidy from the Cabinet Office of Japan given to the Integrative Systems Biology Unit (Marquez-Lago lab), OIST.

**Competing Interests Statement**

The authors have declared that no competing interests exist.

**Manuscript Figures**

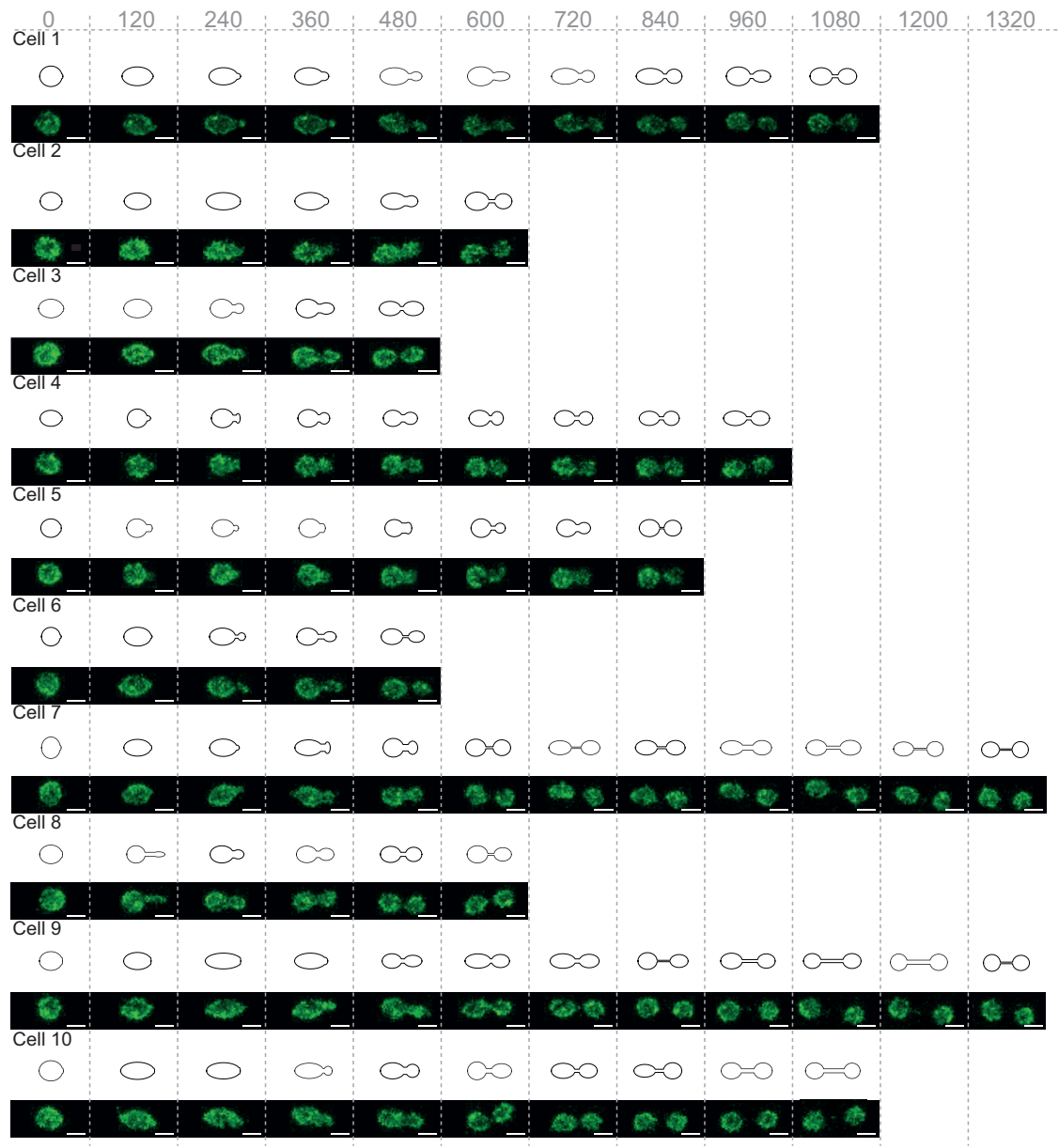

**Fig. 1: Measured cell geometries and their in silico representation.** The geometries of ten yeast nuclei during anaphase as seen through confocal microscopes and their representation in computer simulations. Nuclear pores were visualised using a NUP49-GFP fusion. Whereas plasmids were visualised by the expression of LacI-GFP and their subsequent binding to *lacO* sites. Scale bars are 2μm in length. Geometries were measured in 30 second increments, but are shown in 120 second increments due to space constraints.

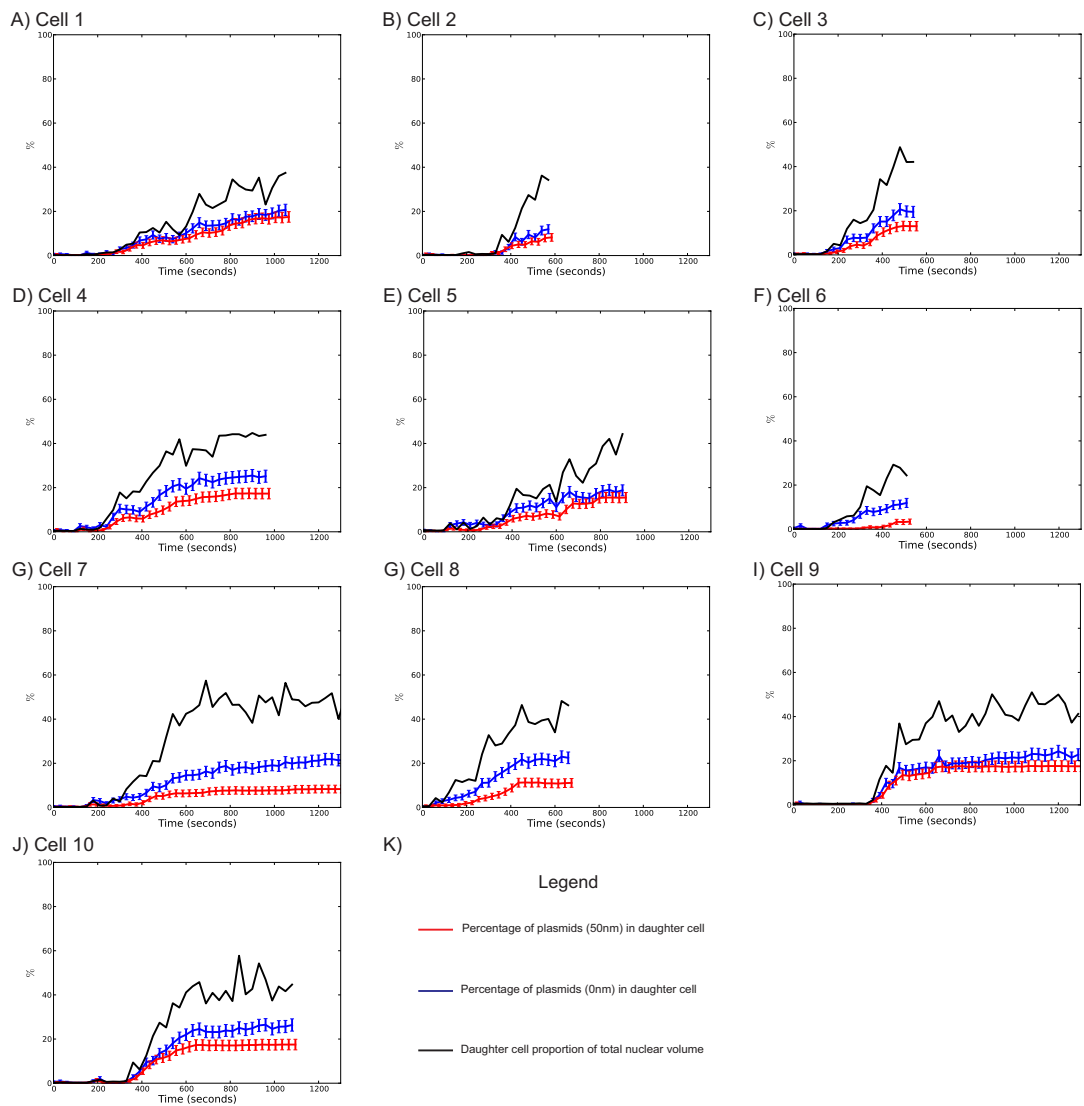

**Fig. 2: Simulated percentage of plasmids in the daughter cell through the course of anaphase for 10 different cell geometries.** Black lines show the proportion of the total nucleus contained within the daughter. Blue lines show the percentage of point particles and contained within the daughter through time. Red lines represent the percentage of 50nm spheres contained within the daughter through time. Error bars show 95% confidence intervals calculated assuming binomial samples at every time point.

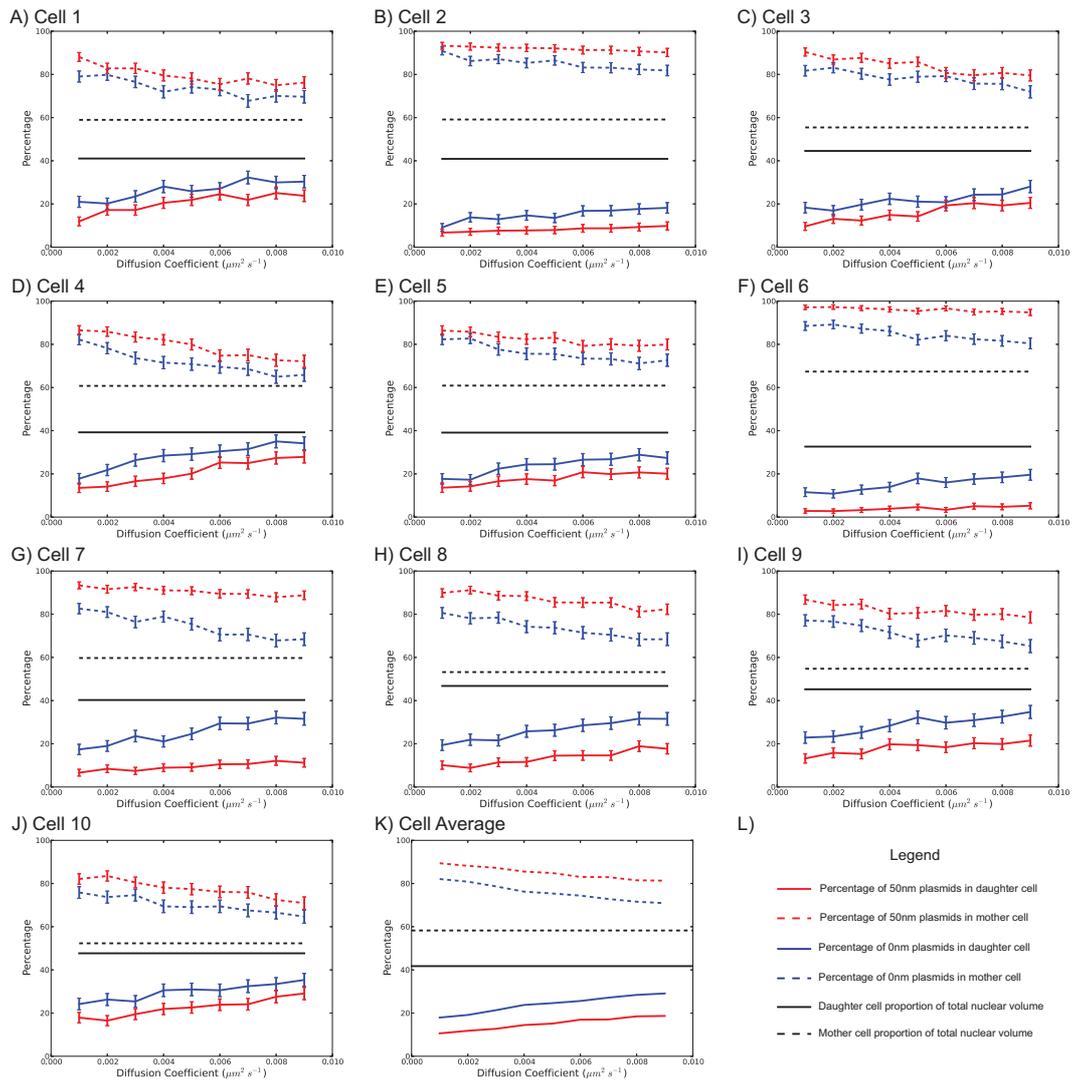

**Fig. 3: Simulated final percentage of plasmids against plasmid diffusion coefficients.** Point particles and spheres with 50 nm radii are shown by blue and red lines, respectively. Black lines indicate the proportion of the nuclear volume contained by the daughter and mother immediately preceding karyofission. Solid and dashed lines represent the final proportions within the daughter and mother, respectively. Subplots a-j represent the results for 10 different cell geometries. Subplot k shows the mean of subplots a-j. Error bars show 95% confidence intervals, and were calculated assuming a binomial distribution.

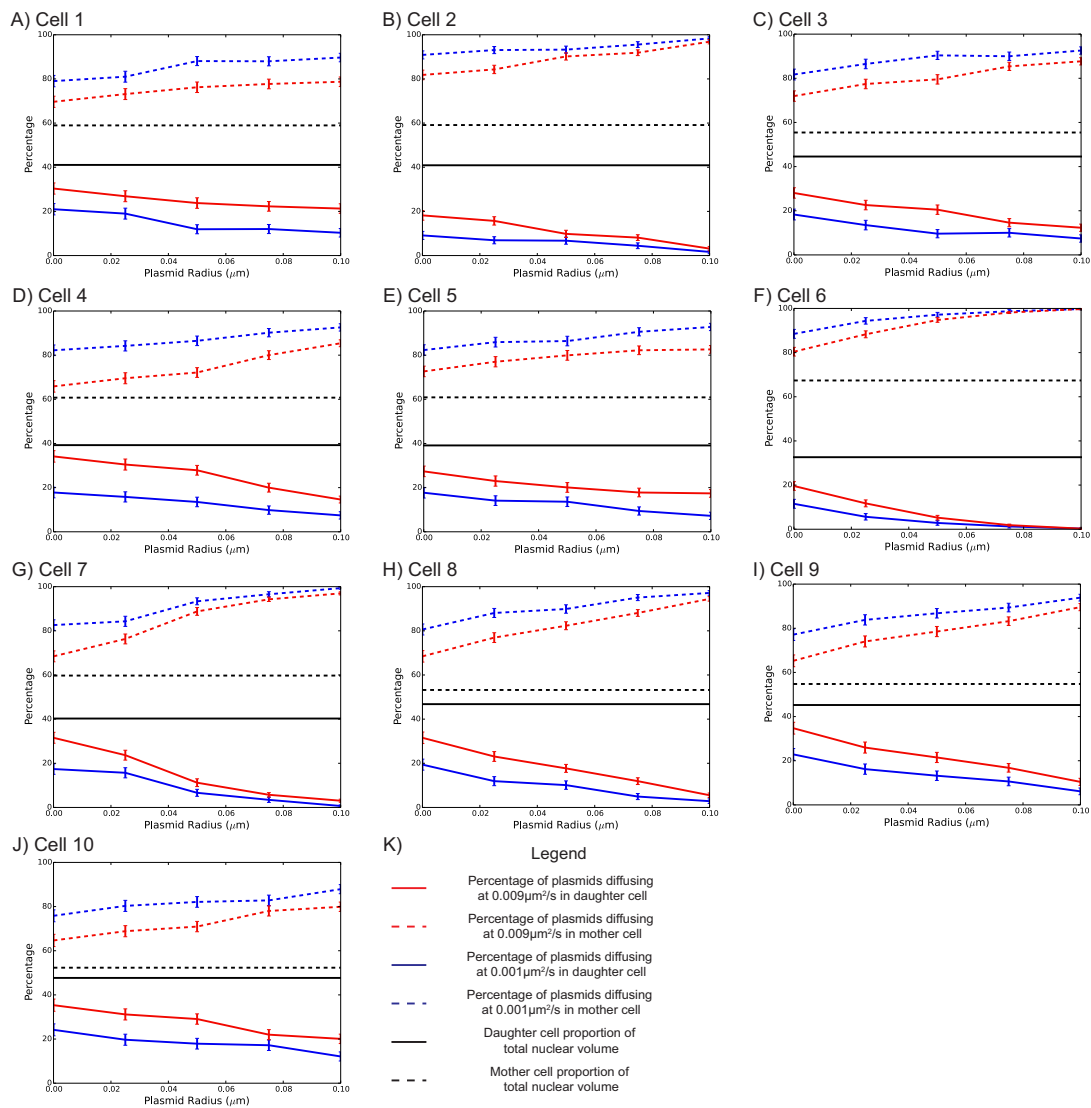

**Fig. 4: Simulated final percentage of plasmids against plasmid radius, assuming spherical plasmids of different sizes.** Diffusion constants of 0.001 $\mu m^2/s$ and 0.009 $\mu m^2/s$ are shown by blue and red lines, respectively. Black lines indicate the proportion of the nuclear volume contained by the daughter and mother immediately preceding karyofission. Solid and dashed lines represent the final proportions within the daughter and mother, respectively. Subplots a-j represent the results for 10 different cell geometries. Subplot k shows the mean of subplots a-j. Error bars show 95% confidence intervals, and were calculated assuming a binomial distribution.

**Supplementary Material**

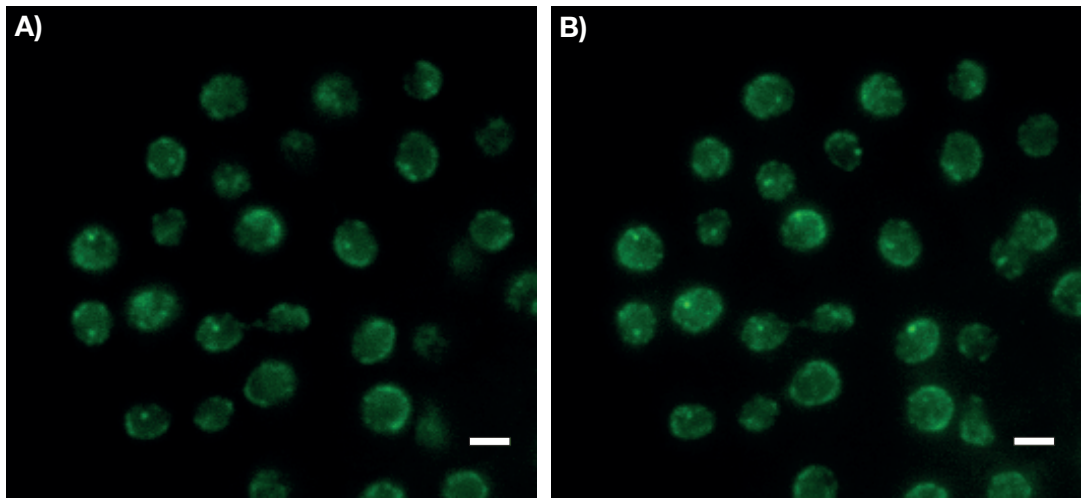

**Fig. S1: Example of PS1 geometries demonstrating limitations in the single colour experimental system.** Example of single time points of GA1320 cells. Given both the nuclear pore and plasmid are marked with GFP, direct plasmid diffusion measurements can become difficult. A) A single Z-stack taken from a stack of 6 separated by 0.5um. B) Maximum intensity projection of all 6 Z-Stack slices.

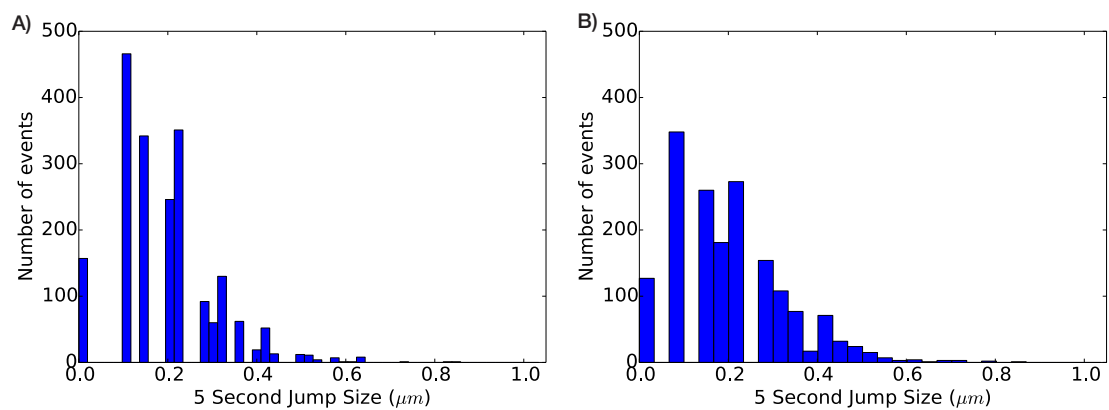

**Fig. S2: Diffusion Jump Size Distribution.** Histograms showing the frequency of jump sizes of plasmids sampled at 5 second increments from 24 distinct cells in interphase (subplot a) and anaphase (subplot b). Histogram bin sizes were calculated according to the Freedman-Diaconis method [27]. Gaps in the histogram occur due to the plasmid jumping over discrete pixels. Owing to the difficulty in accurately estimating diffusion coefficients from this data, our study investigated a range of diffusion rates in simulations.

**Table S3: Specific diffusion coefficients calculated per individual cell according to cell-phase.**

Individual diffusion constants estimated per cell, chosen according to their stage in the cell cycle. Coordinates of the plasmid were tracked every 5 seconds in ImageJ. For each cell, a distribution of 5-second jump sizes was calculated in a 2D plane. Assuming Brownian motion, we have that the diffusion coefficient, D, is then given by: $D = \frac{\langle x^2 \rangle}{2d\Delta t}$, where $\langle x^2 \rangle$ is the average jump size, $\Delta t$ is the time interval (5 seconds) and d is the dimensionality (2, since we only track the movement in a plane). Particle jumps near nuclear membranes were ignored, since the input for simulations should be a free diffusion coefficient. This leads to the following diffusion coefficients, all in $\mu m^2/s$.

| Interphase | Anaphase |
|---|---|
| 0.0016 | 0.0035 |
| 0.0023 | 0.0036 |
| 0.0035 | 0.0020 |
| 0.0022 | 0.0010 |
| 0.0043 | 0.0032 |
| 0.0014 | 0.0040 |
| 0.0026 | 0.0027 |
| 0.0029 | 0.0034 |
| 0.0028 | 0.0025 |
| 0.0021 | 0.0026 |
| 0.0022 | 0.0031 |
| 0.0027 | 0.0022 |
| 0.0022 | 0.0024 |
| 0.0020 | 0.0043 |
| 0.0016 | 0.0022 |
| 0.0016 | 0.0045 |
| 0.0026 | 0.0046 |
| 0.0022 | 0.0026 |
| 0.0017 | 0.0037 |
| 0.0030 | 0.0037 |
| 0.0029 | 0.0023 |
| 0.0043 | 0.0024 |
| 0.0031 | 0.0037 |
| 0.0028 | 0.0031 |

The average of all such cells is 0.0027 $\mu m^2/s$. The average of interphase cells is 0.0025 $\mu m^2/s$; the average of anaphase cells is 0.0031 $\mu m^2/s$.

**Movie S4: Sample simulation output.** A sample of simulation output of a single plasmid, modelled as a point particle, diffusing at 0.0027 $\mu m^2$/s, in the geometry represented by Cell 8 in Figure 1.

**Text S5: A brief overview of the simulation methodology, with references to more detailed studies of the correctness of the procedure used.**

The methodology is described in detail in [1]. We provide a brief overview here, and refer readers there for a full discussion with validation of the technique.

A general yeast nuclear geometry was represented by two oblate spheroids connected by a cylindrical bridge. Each oblate spheroid was truncated in such a way such that the cylindrical bridge would adjoin it without creating any gaps. Thus, we assume a rotationally symmetric domain for the nuclear geometry. The size of each oblate spheroid could thus be described by two numbers describing the major and minor axes. The cylinder adjoining them could be described by the cylindrical length and diameter. The figure below illustrates a sample geometry projected onto a 2D plane.

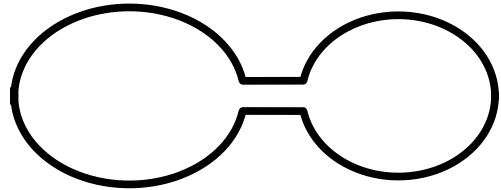

The parameters describing the geometry were measured by confocal microscopy at 30 second intervals. In order to interpolate between the geometries smoothly, the parameters were linearly interpolated between the 30 second time steps. The geometries were constructed in such a way so that the centre of mass of the geometry was at the origin at every time step.

At every time step, a particle was propagated according to a Gaussian kernel. In case the particle would overlap with the boundary, a vector connecting the particle position to the closest point on the boundary was constructed. The particle was then moved in the direction of this vector such that the particle would be completely contained within the boundary once more.

The midpoint of the bridge was taken to be the demarcation zone between the mother and the daughter.

[1] Ghosh, A., Marquez-Lago, T.T., 2014, Simulating Stochastic Reaction-Diffusion Systems on and within Moving Boundaries  arXiv:1410.8596

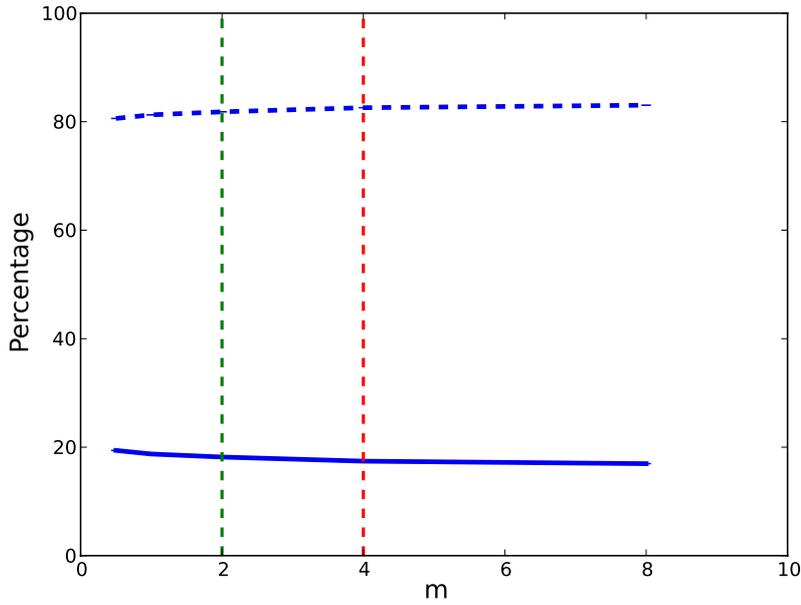

**Fig. S6: Effect of simulation time step selection:** The proportion of plasmids segregating to the mother (dotted line) and daughter (solid line) as a function of simulation time step. Cell 1 was used for cell-geometries. Time steps are controlled by varying a parameter *m*, using the expression: $\delta t = \frac{s_{res}^2}{mD}$, with $s_{res} = 0.025\mu m$ and $D = 0.001 \mu m^2/s$. Simulations were conducted 5000 times. Error bars for the simulations are omitted, since they are not visible beyond the line thickness of the plot. Dashed green lines show a conventional choice of time step for simulations with a static boundary. Dashed red lines represent the values we use in the current study.

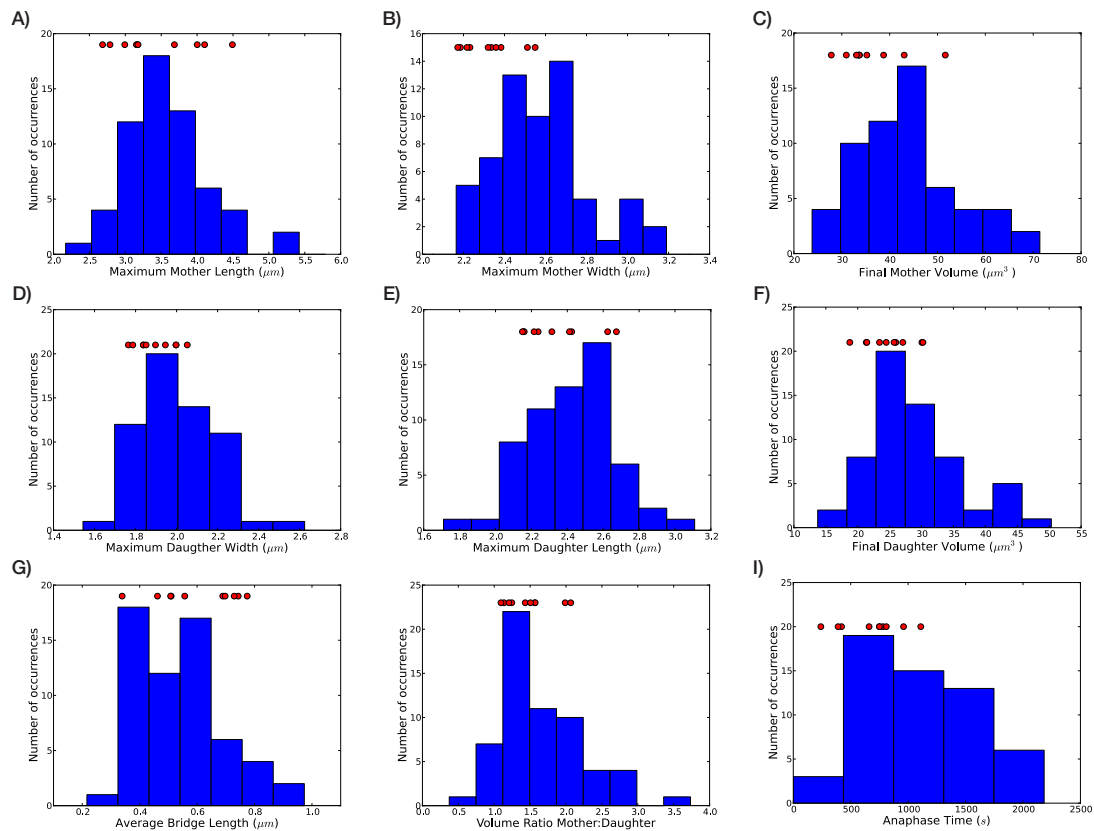

**Fig. S7: Image Geometry Comparison:** A series of metrics used to examine the representative nature of geometries used in the simulations. The initial ten nuclear geometries are represented by red dots in each frame. An additional 50 cells where measured and the distribution for each nuclear metric is represented with a histogram representing: A) The maximum length of the mother nucleus; B) The maximum width of the mother nucleus; C) The final volume of the mother nucleus; D) The maximum length of the daughter nucleus; E) The maximum width of the daughter nucleus; F) The final volume of the daughter nucleus; G) The average bridge width between the dividing nuclei; H) The volume ratio between divided nuclei; I) The anaphase time from the formation of the daughter bud to the clear separation of the divided nuclei.

**S8 Geometry Measurements:** A spread sheet containing the exact measurements for each of the geometries used in the simulations.